\documentclass[pra,aps,twocolumn,showpacs,10pt,floatfix]{revtex4-1}
\usepackage{amsmath,amssymb,graphicx,bm,color,float,ulem}
\usepackage{amsfonts,amsthm}

\begin{document}

\title{Teleportation of a quantum particle in a potential via quantum Zeno dynamics}

\author{Miguel A. Porras}
\address{Grupo de Sistemas Complejos, ETSIME, Universidad Politécnica de Madrid, Rios Rosas 21, 28003 Madrid, Spain}
\author{Miguel Casado-\'Alvaro}
\address{Departamento de Energía y Combustibles, ETSIME, Universidad Politécnica de Madrid, Rios Rosas 21, 28003 Madrid, Spain}
\author{Isabel Gonzalo}
\address{Departamento de Óptica, Facultad de Ciencias Físicas, Universidad Complutense de Madrid,
Ciudad Universitaria s/n, 28040 Madrid, Spain}

\begin{abstract}
We report on the possibility of teleportation of a quantum particle, a distinctly different phenomenon from the teleportation of a quantum state through entanglement. With the first meaning, teleportation is theoretically possible by placing the particle initially at rest (with a certain uncertainty) out of any equilibrium point of a potential well or barrier and by frequently monitoring whether the particle remains at rest. This quantum Zeno dynamics inhibits acceleration, and features disappearance from the classical turning point and appearance in other turning point, if there is any other, with a probability that approaches unity by increasing the frequency of the measurements. This phenomenon has all the ingredients attributed in science fiction to teleportation: The particle is always at rest, cannot be found in the path between the two turning points, and saves travel time. We discuss the feasibility, in principle, of teleportation of electrons, protons and other particles, and conclude its increasing impracticability as the particle gets heavier.
\end{abstract}


\maketitle

\section{Introduction}

Since the nineties of last century, quantum teleportation is a technique of transmitting quantum information from a emitter to a receiver over a distance. It requires quantum entanglement, and therefore several particles \cite{bennett,zeilinger,pirandola}. The possibility of teleportation with the meaning prior to the above achievements, i.e., the teleportation of a single object, has deserved very little attention. Without a scientific basis, the hypothetical teleportation of a particle relies on science fiction literature since the 19th century, on popular culture, and since the cinema, on sci-fi movies. It would involve 1) disappearance of the particle from one location of space, 2) appearance in other distant location, 3) being permanently at rest, and 4) following no path joining the two locations. Teleportation is, also hypothetically, a way to save time on the trip. To our knowledge, only Ref. \cite{wei} discusses the possibility of teleportation in a fractional Schr\"odinger equation, approached by the high-speed limit of relativistic quantum mechanics, and relates teleportation with superconductivity and superfluidity. Also in Ref. \cite{wei}, and more closely related to this paper, the possibly of teleportation in the standard,  nonrelativistic Schr\"odinger equation with measurements of wave function parity is briefly discussed.

Here we report on the possibility of teleportation of a quantum particle with all the above sci-fi characteristics. Teleportation is shown here to result from a peculiar quantum Zeno dynamics (QZD) that has not been explored so far. Quantum Zeno dynamics \cite{facchi1,facchi2,facchi3} must be distinguished from the simpler quantum Zeno effect \cite{misra}. In the quantum Zeno effect, the temporal evolution of a quantum system is interleaved by measurements of whether the state remains the initial one, which results in freezing the evolution in the initial state as the measurements are more frequent. In QZD, measurements ascertain whether the state remains in a set of states, or a multidimensional subspace of the system Hilbert space. The QZD has been studied theoretically \cite{facchi1,facchi2,facchi3}, and has been recently realized in a rubidium Bose-Einstein condensate in a five-level Hilbert space \cite{schafer}.

Beyond a discrete and finite number of states, most of studies of QZD in an infinite-dimensional Hilbert space involve repeated measurements of position which, being necessarily imprecise, monitor whether the particle remains in the region of space, say $\Delta x$, where it is initially found. Reference \cite{facchi4} demonstrates that von Neumann projective measurements of position lead to a unitary evolution confined in the subspace defined by $\Delta x$ with Dirichlet boundary conditions in the limit of continuous measurements, i.e. when the number of measurements $N$ approaches infinite. Different QZD with position measurements have been studied and featured different effects such as inhibition of wave packet spreading of a particle at rest \cite{porras1}, and stopping a freely moving particle (the Zeno arrow) in a Schr\"odinger cat state \cite{porras2}.

More recently, Porras {\it et al} \cite{porras3} have analyzed the conjugate QZD involving repeated measurements of momentum. Specifically, frequently monitoring whether the momentum of a particle directed towards a potential barrier remains positive results in freezing the momentum direction and hence in an Zeno-assisted quantum tunneling, with tunneling probability approaching unity as the number of measurements increases.

The QZD leading to teleportation is that of a particle initially at rest, that is, $\langle v\rangle=0$, 
out of an equilibrium point in a potential, hence  experiencing a force at a classical turning point, whose state of rest with uncertainty $\Delta v$ is frequently monitored. The resulting QZD tends to keep the particle at rest, i.e., the quantum state in the subspace of velocities $|v|<\Delta v$, with higher probability with growing number of measurements, and despite the force acting on the particle.

We find that the way of remaining at rest is at least a bit striking (see video in \cite{video}). The particle disappears from its initial location while appearing on other classical turning point, if there is any other, exhibiting all the above mentioned characteristics of teleportation. When the number of measurements $N$ increases, the probability of teleportation approaches unity in a teleportation time that is independent of $N$. The teleportation time is proportional to the mass and $\Delta v$, but shorter as the force is stronger, and can be made much shorter than the time taken by the particle to travel up to the other turning point (see also video in \cite{video}). 

We demonstrate the absence of probability flux in the path between the two turning points, which amounts a violation of the equation of continuity for the probability in this QZD, and rigorously supports referring to it as teleportation. Eventually, we examine how the mass affects the teleportation probability with different schemes of measurements. The conclusion here is that teleportation remains possible, in principle, for heavier particles, but becomes increasingly difficult, and completely impracticable for macroscopic objects.

\section{Zeno dynamics of a particle in a potential monitoring being at rest}

\begin{figure}[t]
\centering
  \includegraphics*[width=8cm]{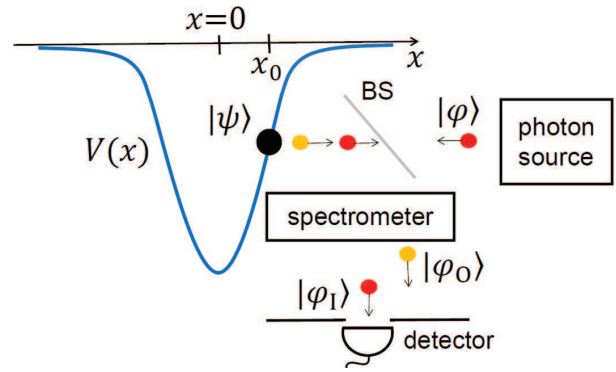}
   \caption{\label{fig1} The backscattered photon by the particle experiences a small Doppler shift (red photon) if the particle is at rest ($|v|<\Delta v$),
   or a large shift (e.g., orange photon) if the particle is not at rest, in which case measurements cease.}
\end{figure}

We start with Schr\"odinger equation for a particle of mass $m$ in a potential $V$, $i \hbar \partial |\psi\rangle/\partial t = \hat H |\psi\rangle = (\hat p^2/2m)|\psi\rangle + \hat V|\psi\rangle$ written for the wave function $\psi(x,t)=\langle x|\psi\rangle$ in one dimension and in atomic units 
($\hbar =1$, $m_e=1$)
as
\begin{equation}\label{SCH}
i\frac{\partial \psi}{\partial t} = -\frac{ 1}{2m}\frac{\partial^2\psi}{\partial x^2} +  V(x)\psi\,.
\end{equation}
For simplicity we will take symmetric potentials about $x=0$, as in Fig. \ref{fig1}. The particle is initially at rest and localized about a position $x_0=\langle x\rangle$ different from an equilibrium point, i.e., at a classical turning point experiencing a force, and such that there is only one other turning point $-x_0$. The announced results hold for barrier or well potentials indifferently, as well as for non-symmetric potentials with more turning points, but their consideration only distracts from the essentials. Being initially at rest means $\langle v\rangle=0$ with a certain uncertainty $\Delta v$, i.e., $|v|\le \Delta v$. Considering initial Gaussian-like wave functions $\psi(x,0)=\langle x|\psi_0\rangle$ and $\hat \psi(p,0)=\langle p|\psi_0\rangle$ of half Gaussian widths ($1/e^2$ decay in probability) $\Delta x$ and $\Delta p$, such minimal wave packets verify $\Delta x \Delta p= 2$, hence, given $\Delta v$, $\Delta x=2/m\Delta v$, but this is only a choice to fix ideas.

We next consider the following QZD. In a given time interval $T$, the particle wave function is left to evolve in
small time intervals $\Delta t = T/N$ according to Eq. (\ref{SCH}), interleaved with measurements of whether the particle remains at rest with the same uncertainty $\Delta v$ about zero as initially. The number of measurements of velocity is then $N$. They are modelled as projections of the state onto the subspace of velocities $|v|<\Delta v$, or equivalently onto the subspace of momenta $|p|<m\Delta v$. We consider only positive outcomes ($|v|<\Delta v$) in all the measurements, i.e., the measurements are selective, and wonder about the probability $P_N^{(S)}$ that all measurements are positive. Upon a negative outcome the measurements are terminated. 

Figure \ref{fig1} shows a possible implementation of measurements based on the Doppler effect of photons interacting with the particle. A backscattered photon may be Doppler shifted within $\Delta \lambda$ corresponding to velocities in $\Delta v$ (small Doppler shift represented by red photons in the figure)
with state $|\varphi_I\rangle$, or outside $\Delta \lambda$ (large Doppler shift represented by orange photons, for example) with state $|\varphi_O\rangle$. After the interaction the particle-photon entangled state is $|\psi_I\rangle |\varphi_I\rangle +|\psi_O\rangle |\varphi_O\rangle$, where
\begin{equation}
|\psi_I\rangle = \Pi |\psi\rangle =\int_{|p|<m\Delta v}\!\!\!\!\!dp|p\rangle \langle p|\psi\rangle  =\int_{|p|<m\Delta v} \!\!\!\!\!dp\hat\psi(p)|p\rangle  ,
\end{equation}
and
\begin{equation}
|\psi_O\rangle =(1- \Pi) |\psi\rangle =\int_{|p|>m\Delta v}\!\!\!\!\!dp|p\rangle \langle p|\psi\rangle  =\int_{|p|>m\Delta v} \!\!\!\!\!dp\hat\psi(p)|p\rangle .
\end{equation}
The detector measures if the wavelength is within $\Delta\lambda$ reducing the state of the particle to $|\psi_I\rangle /\sqrt{\langle \psi_I|\psi_I\rangle}$, or outside, in which case the experiment is over. 

\begin{figure*}
\includegraphics*[height=4.2cm]{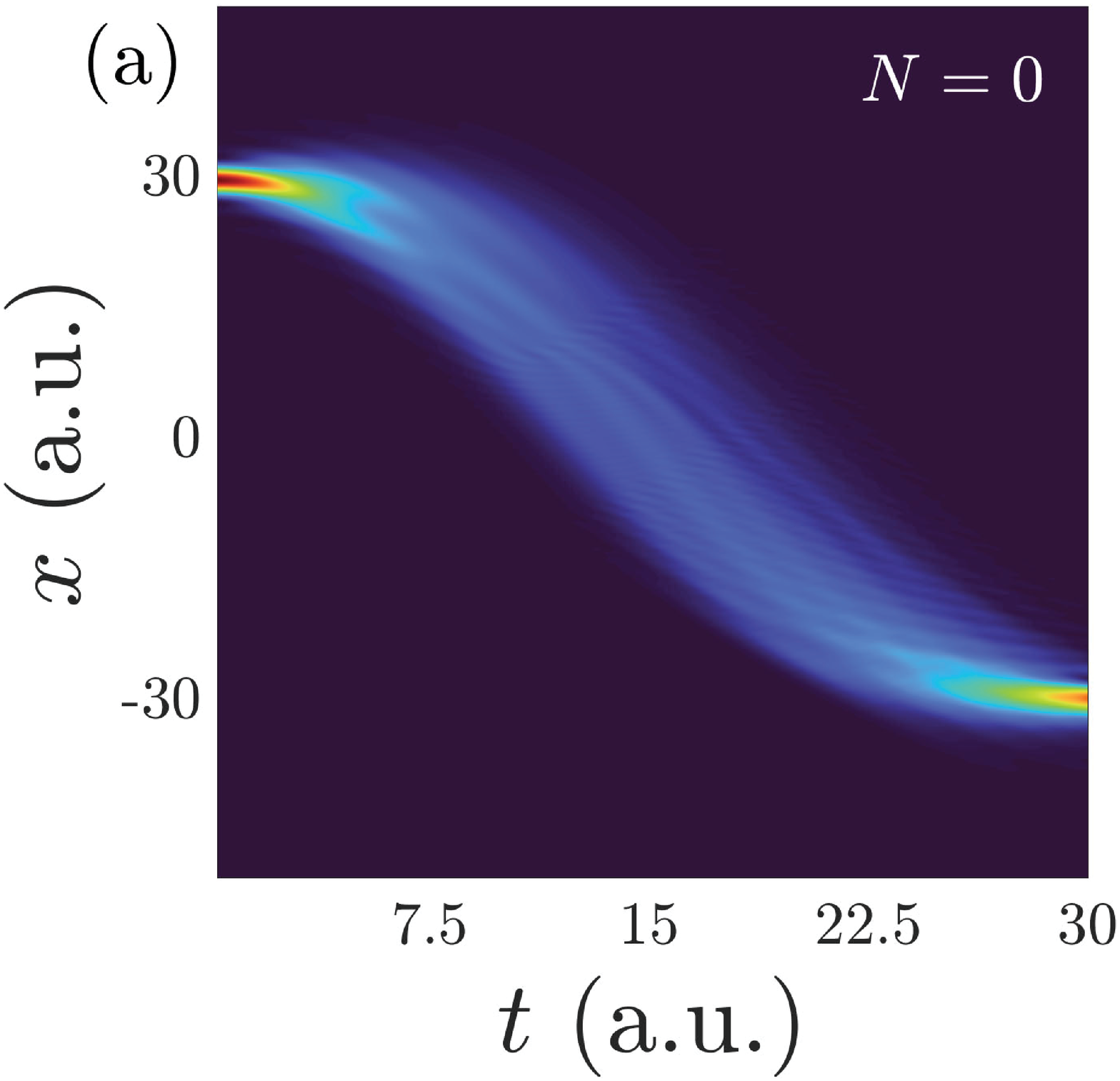}\includegraphics*[height=4.2cm]{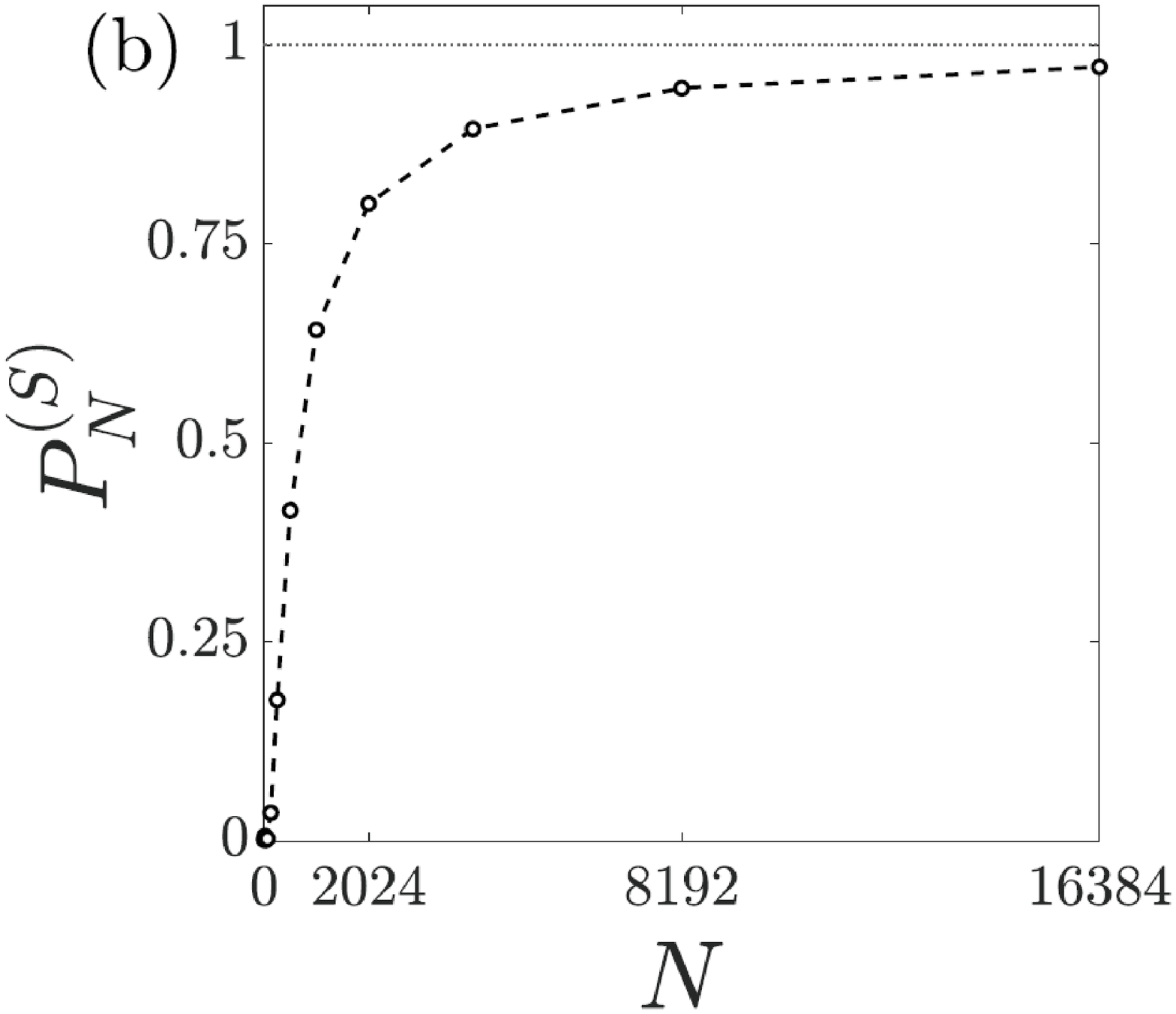}\includegraphics*[height=4.2cm]{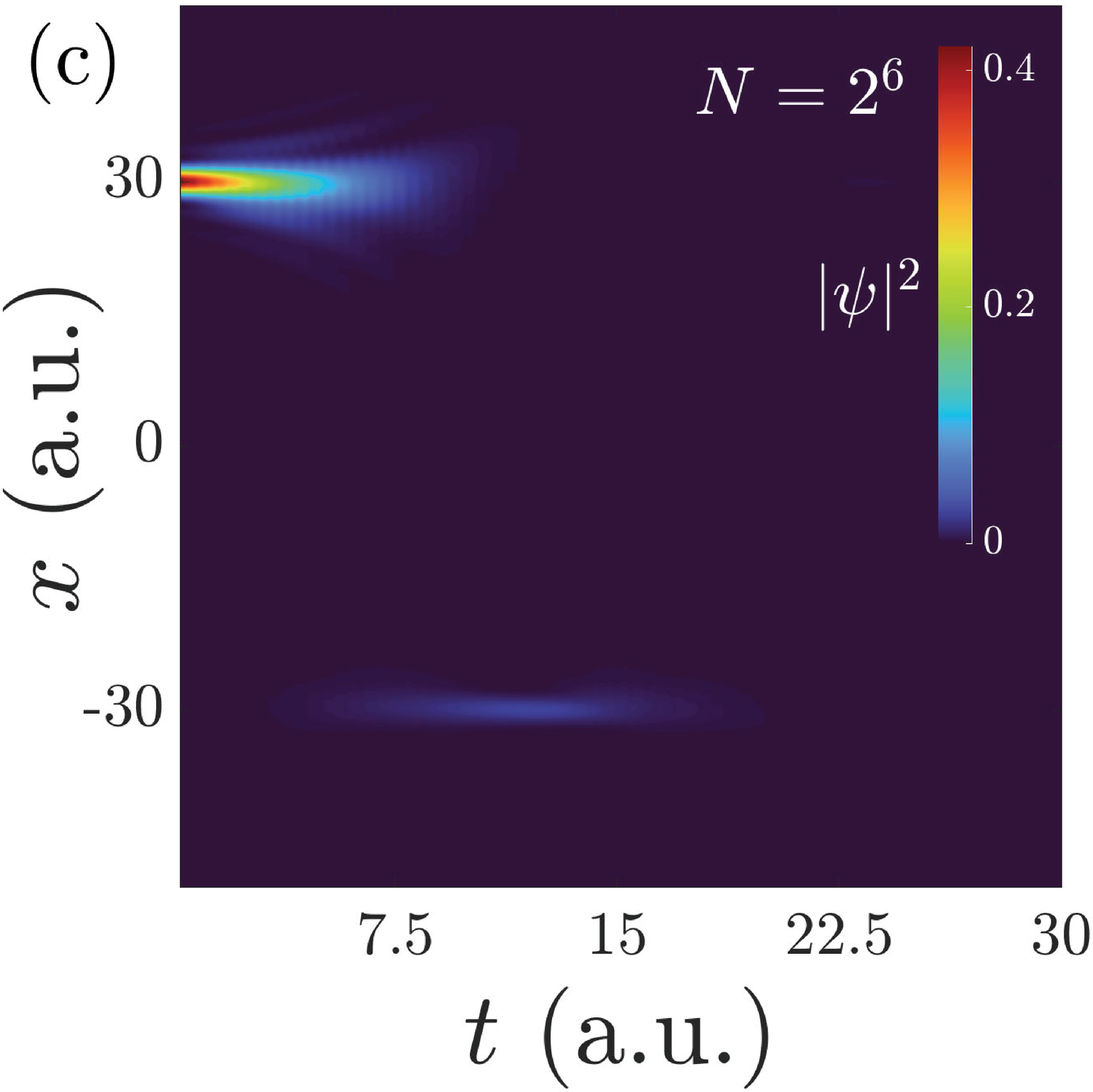}\includegraphics*[height=4.2cm]{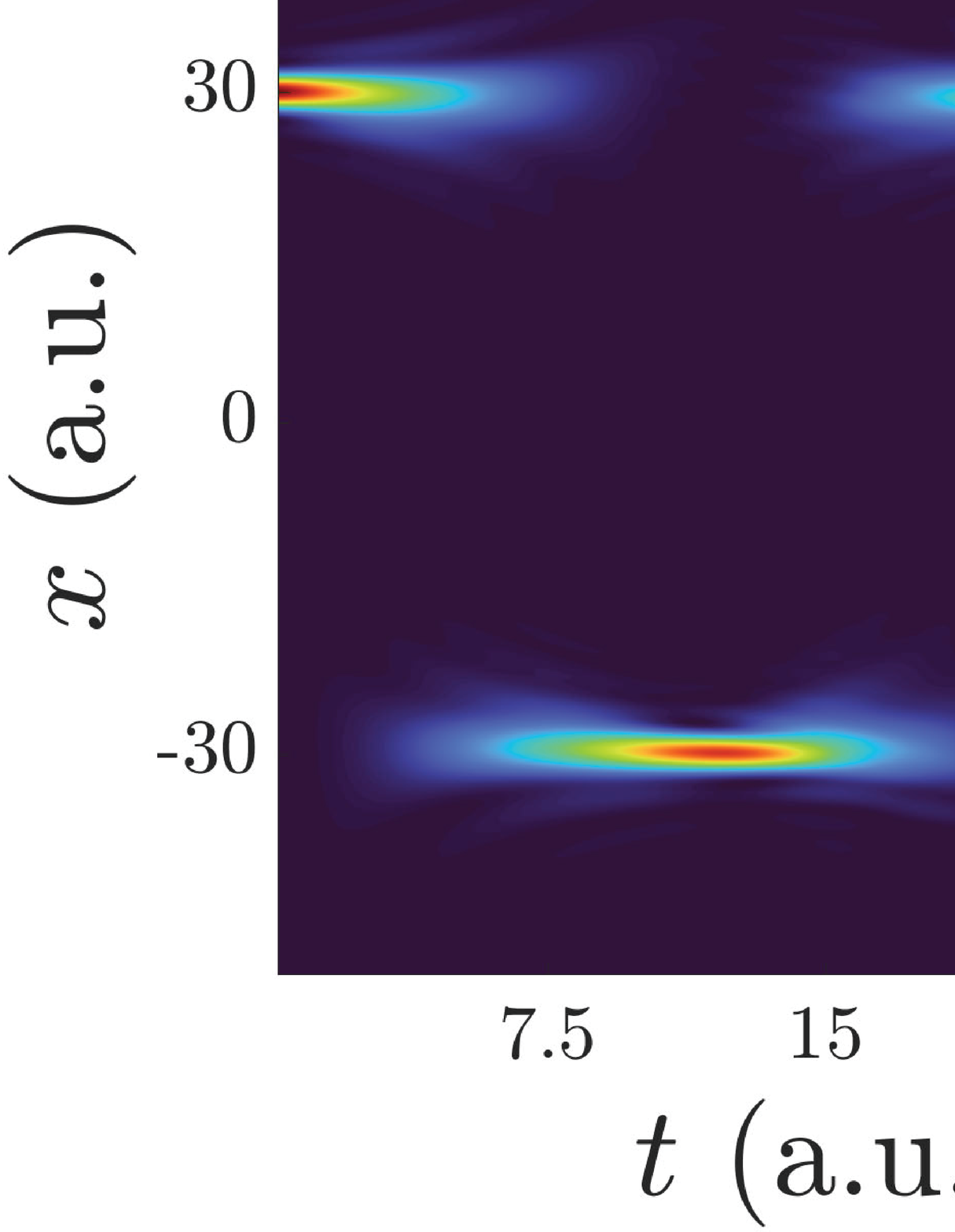}
\caption{\label{fig2} (a),(c) and (d): Contour plots of the probability density 
as a function of time for a particle of mass $m=1$ initially at the state  $\psi_(x,0)=(1/\pi)^{1/4}\exp[-(x-x_0)^2/2]$, $x_0=30$ in the potential well $V(x)=-V_0\exp(-x^2/x_p^2)$, $x_p=30$ and $V_0=10$.  All quantities are in atomic units. (a) Without any measurement. (c) and (d) With the indicated number of measurements. (b) Probability that the particle remains at rest with velocity $|v|\le \Delta v=\sqrt{2}$ at the final time $T=30$ as a function of number of measurements $N$.}
\end{figure*}

The QZD with these selective measurements can then be symbolized as
\begin{equation}\label{NORM}
|\phi_N\rangle =\frac{1}{\sqrt{P_N}} \Pi e^{-i\hat H\Delta t}\!\!\dots \frac{1}{\sqrt{P_2}} \Pi e^{-i\hat H\Delta t} \frac{1}{\sqrt{P_1}} \Pi e^{-i\hat H\Delta t} |\psi_0\rangle\,,
\end{equation}
where $P_n$ is the probability that the velocity is in $[-\Delta v,\Delta v]$ (momentum in $[-m\Delta v,m\Delta v]$) at the $n$ intermediate measurement, so that the factors $1/\sqrt{P_n}$ normalize the state after each projective measurement. After the $N$ measurement at time $T$, the probability that all outcomes were positive is the product $P_N^{(S)} = P_1 P_2 \dots P_N$. To obtain $P_N^{(S)}$, however, it is traditional in QZD, and faster computationally, to evaluate the unnormalized state
\begin{equation}\label{UNNORM}
|\psi_N\rangle =\Pi e^{-i\hat H\Delta t}\stackrel{^{\mbox{\small $N-3$ times}}}{\dots\dots} \Pi e^{-i\hat H\Delta t} \Pi e^{-i\hat H\Delta t} |\psi_0\rangle\,,
\end{equation} 
which is obviously related to the normalized state $|\phi_N\rangle$ in (\ref{NORM}) by $|\psi_N\rangle =\sqrt{P_1P_2\dots P_N}\,|\phi_N\rangle$, and whose norm, 
\begin{equation}\label{SELECTIVE}
\langle\psi_N|\psi_N\rangle =P_1P_2\dots P_N =P_N^{(S)},
\end{equation}
yields directly the probability $P_N^{(S)}$ that all outcomes were positive. In all figures, the unnormalized state at any intermediate step $n$ is represented since its norm informs us about the probability that the velocity remains in $|v|<\Delta v$ at the intermediate step $n$.

We have implemented the above QZD in a numerical code, where unitary evolutions in
$\Delta t$ are performed using a split-step Fourier method to solve Schr\"odinger equation, and measurements by truncating the wave function in momentum representation outside $[-m\Delta v,m\Delta v]$. For a potential well of width $x_p$, the initial location is far from the equilibrium point $x=0$ by setting $x_0\sim x_p$. Also, the uncertainty $\Delta v$ is chosen such that $\Delta x =2/\Delta p=2/m\Delta v$ is much smaller than $x_p$ so that the probability of being located at the other turning point is zero to all practical purposes.

\begin{figure*}
\includegraphics*[height=9cm]{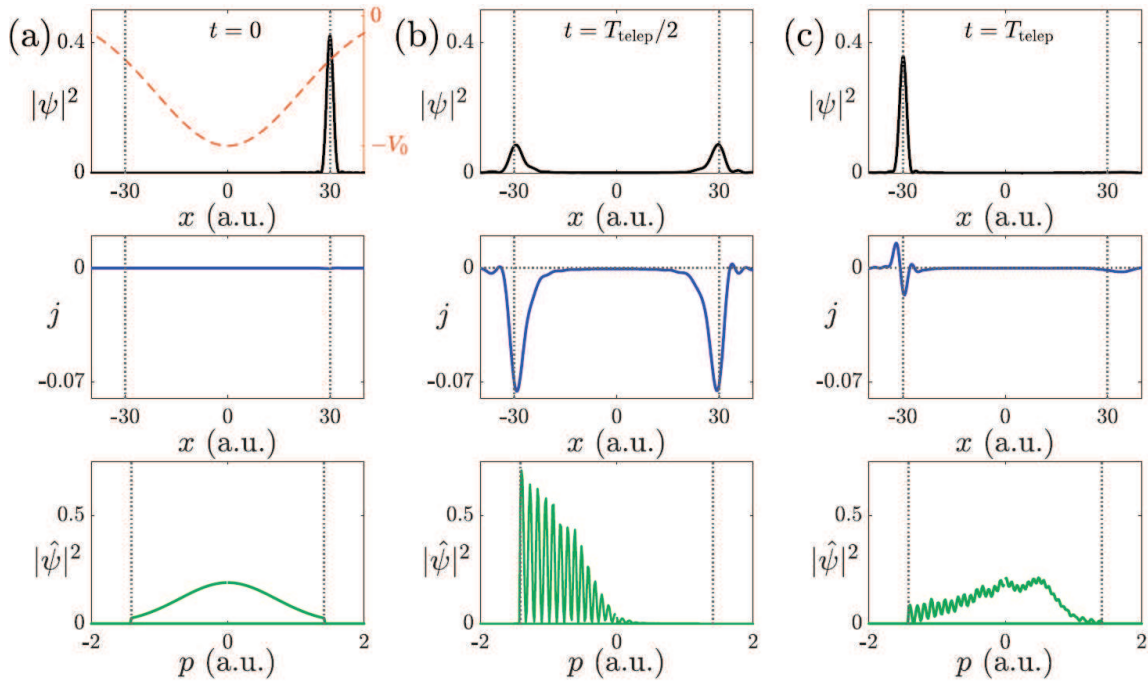}
\caption{\label{fig3} For the same potential well and initial wave function as in Fig. \ref{fig2}, probability density (first row), probability current (second row), and momentum probability density (third row), at (a) t=0, (b) $t=T_{\rm telep}/2$ and (c) $t=T_{\rm telep}=11.533$ a.u. The number of measurements is $N=2^{10}$ and the final probability of finding the particle at $-x_0$ is $P_N^{(s)}=0.93$. The potential well is depicted as a dashed curve in (a, top).}
\end{figure*}

Typical results for $m=1$ (an electron) are shown in Fig. \ref{fig2}. Without measurements, the wave function oscillates in a potential well, as seen in Fig. \ref{fig2}(a) [the potential well is represented in Fig. \ref{fig3}(a)]. With measurements of whether the particle remains at rest ($|v|$ in $\Delta v$), the particle tends to remain at rest with probability $P_N^{(s)}$ approaching unity as the number of measurements $N$ within the time $T$ increases, as seen in Fig. \ref{fig2}(b). This  QZD then features freezing the velocity about zero against the permanent the force acting on it. Figures \ref{fig2} (c) and (d) show the striking way in which the particle stay still: The particle ``finds" the other turning point where it can stay still, $-x_0$, and appears and disappears periodically in $-x_0$ and $x_0$ with a period substantially independent of $N$ and ostensibly smaller than the natural period of oscillations.

\section{Teleportation}

Given the novelty of this phenomenon, from now on we set $T$ equal to the time that the particle takes to teleport from one to another turning point, or teleportation time $T_{\rm telep}$, defined as the time at which the probability density at $-x_0$ is maximum. Since no appreciable norm is left at $T_{\rm telep}$ at $x_0$, the probability $P_N^{(S)}$ is the probability of teleportation. A closer look at examples allows us to identify the mechanism of teleportation and to estimate $T_{\rm telep}$. 

\subsection{Teleportation mechanism}

Figure \ref{fig3} shows the probability densities in position and momentum spaces at the selected times (a) $t=0$, (b) $T_{\rm telep}/2$ and (c) $T_{\rm telep}$ a.u. for the same initial state and potential well as in Fig. \ref{fig2}. The number of measurements $N=2^{10}$ is high, hence the probability of teleportation $P_N^{(s)}=0.93$ is also high. A video of the whole dynamics can be found in the supplemental material \cite{video}. At no time does $|\psi(x,t)|^2$ take appreciable values in the path between $x_0$ and $-x_0$ (first row in Fig. \ref{fig3}). We discard a significant  probability current or flux $j(x,t)=(1/m)\mbox{Im}\{\psi^\star \partial\psi/\partial x\}$ between $x_0$ to $-x_0$ (second row). This point is  further discussed below. 

The mechanism of teleportation can be understood from the dynamics of the wave function in momentum representation. We observe that the repeated measurements create a sharp potential well in momentum space extending from $-\Delta p=-m \Delta v$ to $\Delta p = m \Delta v$ (third row). In the limit of continuous measurements, the well would become infinitely high, with the momentum wave function $\hat \psi(p)$ satisfying Dirichlet boundary conditions at  at $-\Delta p = -m \Delta v$ and $\Delta p =m \Delta v$, as in \cite{facchi4} for position measurements. With finite measurements the reflections produce norm losses that are smaller as $N$ increases. We also observe that the reflection is about the time $T_{\rm telep}/2$ at which the particle is located at $x_0$ and $-x_0$ with approximately equal probabilities, and that a complete round trip to the original location of the momentum wave function coincides with $T_{\rm telep}$ at which the particle is only located at $-x_0$.

Thus, writing the initial wave function as $\psi(x,0)=\psi(x-x_0)$, or $\hat \psi(p,0)=\hat \psi(p)e^{-ix_0p}$, the wave function at the reflection time is a superposition of left and right moving momentum wave functions $\alpha \hat \psi(p) e^{-ix_0 p}+\beta \hat \psi(p) e^{ix_0 p}$, leading to the interference fringes, with decreasing $\alpha$ and increasing $\beta$ as the reflection takes place, or equivalently, position wave function $\alpha \psi(x-x_0) + \beta\psi(x+x_0)$ disappearing from $x_0$ and appearing in $-x_0$.

The mean value of momentum in this reflection follows approximately the laws of classical mechanics. Since the particle does not appreciably moves from $x_0$ while it disappears, the force acting on it takes the approximately constant value $F= -V'(x_0)$, where $V'=dV/dx$. The mean momentum as a function of time is then $\langle p\rangle \simeq -V'(x_0) t = -|V'(x_0)| t$, which equated to $-\Delta p$ yields the reflection time as $\Delta p/|V'(x_0)|$. Similarly, from the reflection, the mean momentum changes with time as $\langle p\rangle \simeq -\Delta p -V'(-x_0) t= -\Delta p +|V'(x_0)| t$, which equated to zero yields again $\Delta p/|V'(x_0)|$. The teleportation time is then twice this quantity, i.e.,
\begin{equation}\label{TELEP}
T_{\rm telep} \simeq \frac{2\Delta p}{|V'(x_0)|}= \frac{2m\Delta v}{|V'(x_0)|}\,.
\end{equation}
This expression provides an excellent approximation to the time at which the probability of finding the particle at its original location is zero. Small deviations are due to the slight deviation of the force from $-V'(x_0)$ when the particle slightly moves at $x_0$ or at $-x_0$. The absolute values are introduced by purpose to make the expression valid for any symmetric potential (well or barrier) with the particle initially at any positive or negative location. For a non-symmetric potential the teleportation time would be $m\Delta v/|V'(x_0)|+ m\Delta v/|V'(x_1)|$, where $x_1$ is the second turning point. It is also clear from (\ref{TELEP}) that a force is essential to teleportation. In particular, if the particle is placed at an equilibrium point, teleportation to other equilibrium point requires infinite time, i.e., does not occur at all.

If the well is reversed in sign and becomes a barrier, the above results are almost identical as in Fig. \ref{fig3} and hence are not shown. In particular the teleportation time is also given by (\ref{TELEP}). Thus, if a particle is launched against a barrier with kinetic energy less than the peak value $V_0$, the particle can pass through the barrier with arbitrarily high probability via teleportation-assisted tunneling, by measuring whether the particle remains at rest since the instant of time at which the particle is stopped at the first turning point.

\begin{figure}
\includegraphics*[height=4cm]{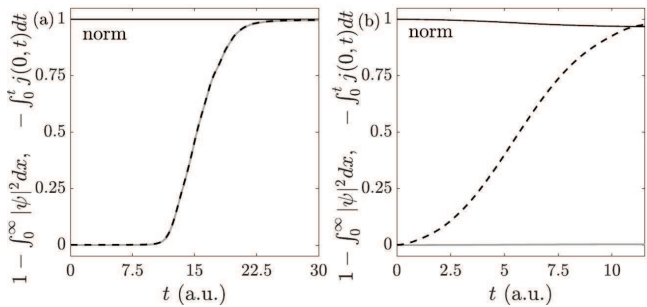}
\caption{\label{fig4} (a) For the example of Fig. \ref{fig2} (a) (no measurements) in the same potential well, the norm is unity (black solid curve), and the decrement of the probability of finding the particle in $[0,\infty)$ (dashed curve) is equal to the outgoing probability through $x=0$ (gray solid curve), as required by the continuity equation. (b) For the same example in Fig. \ref{fig3} with $N=2^{11}$ measurements, the decrement of the probability of finding the particle in $[0,\infty)$ (dashed curve) is not equal to the outgoing probability through $x=0$ (gray solid curve).}
\end{figure}

\subsection{Teleportation as a violation of the continuity equation}

The idea of teleportation, or translation at rest along no path, can more rigorously be associated with the extinction of the probability of presence in a region of space and emergence of presence in another separate region without probability flux through their boundaries. This is a violation of the continuity equation. While the continuity equation is always satisfied in pure Schr\"odinger evolution, it is not necessarily satisfied in a QZD.

In one dimension the continuity equation reads $\partial|\psi|^2/\partial t+ \partial j/\partial x=0$, or in integral form, $(d/dt)\int_a^b |\psi|^2 dx =j(a,t)-j(b,t)$, stating that the variation of the probability in the interval $[a,b]$ per unit time must be due to a probability flux across its boundaries. We take here the relevant interval as $[0,\infty)$ since the particle is initially at the positive $x$-axis. Therefore
\begin{equation}\label{CONTINUITY1}
\frac{d}{dt}\int_0^\infty |\psi(x,t)|^2 dx = j(0,t)-j(0,\infty)= j(0,t)\,.
\end{equation}
Further integration in time and some rearrangement leads to
\begin{equation}\label{CONTINUITY2}
1- \int_0^\infty |\psi(x,t)|^2 dx = - \int_0^t j(0,t) dt\,,
\end{equation}
expressing that, in Schr\"odinger evolution, the decrement of the probability in $[0,\infty)$ [left hand side of (\ref{CONTINUITY2})] is due to the outgoing probability through $x=0$ (right hand side). Without measurements, this is illustrated in Fig.\ref{fig4} (a) with the same example as in Fig. \ref{fig2}(a) (half an oscillation in the potential well). The norm is constant in time (black solid curve), the decrement of the probability of finding the particle in $[0,\infty)$ is complete (dashed curve), and coincides with the outgoing probability through $x=0$ (grey solid curve). Fig. \ref{fig4}(b) shows the same quantities but for the example of Fig. \ref{fig3} with $N=2^{11}$ measurements. In the QZD the norm is almost constant (would be exactly constant in the limit $N\rightarrow\infty$), the decrement of probability of finding the particle in $[0,\infty)$ is also complete, but this decrement is not equal to the outgoing probability through $x=0$ (grey horizontal curve), which is zero at any time. Hence (\ref{CONTINUITY1}) and (\ref{CONTINUITY2}) are not satisfied, supporting that the particle does not follow a path, but literally disappears from $[0,\infty)$. Being the norm almost unity at the end, the particle is located in $(-\infty,0]$.

\section{Teleportation of heavier particles and the classical limit}

The above results simply show the theoretical possibility of teleportation according to standard quantum mechanics. Hereinafter we discuss whether this effect could be possible, also in principle, in hypothetical experiments monitoring rest with typical uncertainties of velocity of particles at rest of different masses, and ultimately, with macroscopic particles.

For an electron $\Delta v=\sqrt{2}$ a.u. ($\sim 10^6$ m/s) and $\Delta x=\sqrt{2}$ a.u. ($\sim 10^{-10}$ m) in the above examples may be reasonable, but not for heavier particles that tend to slow down and to localize. This is reflected by the Boltzmann distribution of velocities $\propto e^{-mv^2/a}$ of one dimensional particles ($a$ would be $a=2k_B{\cal T}$, where $k_B$ is Boltzmann constant and  ${\cal T}$ temperature), of mean value $\langle v\rangle=0$ and width $v_{\rm th} = \sqrt{a/m}$. If these are statistically expected values of classical particles at temperature ${\cal T}$, it is then reasonable to take for our initial pure state $\Delta v =v_{\rm th}$, leading to $\Delta p=\sqrt{am}$ and the increasingly localized wave function with $\Delta x = 2/\sqrt{am}$. Note however that this state does not represent a thermal particle, but simply the statistical uncertainty is assigned to the quantum uncertainty. This assignment yields e.g., $\Delta v\sim 9\times 10^{4}$  m/s and $\Delta x \sim 2\times 10^{-9}$ m for an electron, and $\Delta v\sim 9.2\times 10^{2}$ m/s and $\Delta x \sim 2\times 6^{-11}$ m for a proton at room temperature.

\begin{figure}[t!]
\includegraphics*[height=8cm]{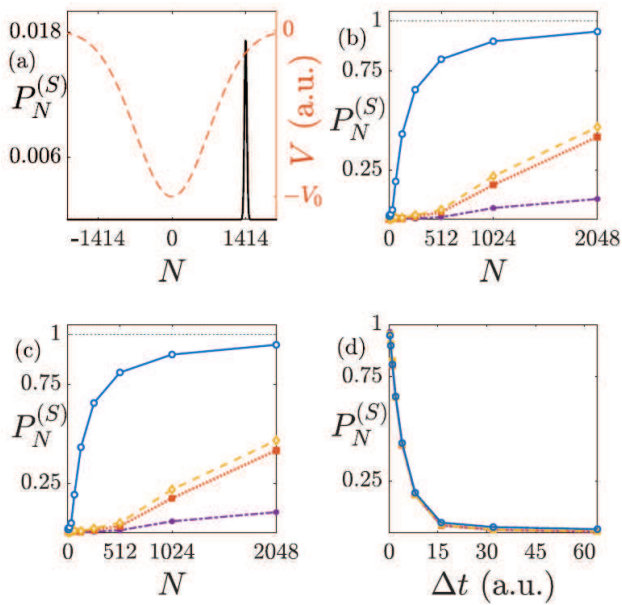}
\caption{\label{fig5} Probability of teleportation in different measurements schemes for four particles: electron ($m=1$, open circles), muon ($m=206.767$, diamonds), pion ($m=273.767$, squares), and proton ($m=1836$, closed circles). (a) shows the initial wave function for the electron $\psi_(x,0)=(2/\pi\Delta x^{2})^{1/4}\exp[-((x-x_0)/\Delta x)^2]$, with $x_0=1414.2$ and $\Delta x = 2/\sqrt(am)$, $a=1.856\times 10^{-3}$ a.u. (the initial wave functions of the other three particles are narrower), and the potential well $V(x)=-V_0\exp(-x^2/x_p^2)$, $x_p=1000$.  All quantities are expressed in atomic units. In (b) the potential depth is set to $V_0=0.4396\sqrt{m/m_e}$, where $m_e=1$ is the electron mass, so that it grows for heavier particles and all of them have the same teleportation time $T_{\rm telep}$. In (c) the potential depth $V_0=0.4396$ is kept constant and therefore the teleportation time increases with mass as in Eq. (\ref{TELEP2}). In (d) teleportation times are the same as in (c), but the interval between measurements, $\Delta t$, is the same for all particles.}
\end{figure}

The teleportation time (\ref{TELEP}) becomes
\begin{equation}\label{TELEP2}
T_{\rm telep} = \frac{\sqrt{4am}}{|V'(x_0)|}\,.
\end{equation}
A particle with the above wave function properties is placed in a potential well, as in Fig. \ref{fig5}(a). For a comparison of the behavior of particles of different masses, the potential has the same shape for all particles, all particles are placed at the same position $x_0$ for the teleportation distance $2x_0$ to be the same. 

We consider four particles of increasing mass. To strictly adhere to the QZD scheme, the teleportation time $T_{\rm telep}$ is the same for the four particles by making the potential depth $V_0$ to grow with mass. The location $x_0$ is also chosen so that $T_{\rm telep}$ is sufficiently long, making not completely unreasonable a high number of measurements; In the example of Fig. \ref{fig5}(b), $N=2^9$ measurements correspond to a measurement each $\Delta t=1$ a.u., about $24$ attoseconds. This figure shows the probability of teleportation as the number of measurements grows for an electron, a muon, a pion and a proton. One could of course prolong the curves to higher number of measurements and all curves will approach unity, but these curves suffice to illustrate the increasing difficulty of teleporting heavier particles.

We may also keep the same well depth for all four particles. Then the teleportation time increases with the square root of mass. With the same conditions as in Fig. \ref{fig5}(a) and (b) except the constant depth $V_0$, Fig. \ref{fig5}(c) shows the probability of teleportation at each teleportation time for the same four particles. Again, heavier particles are teleported with less probability with each given number of measurements. 

We may still argue that in scheme (c), with longer and longer teleportation times and same number of measurements, the time interval $\Delta t$ between them is longer. We then consider, as a third possibility, the probability of teleportation when $\Delta t$ is kept constant for all four particles along the respective teleportation times. Fig. \ref{fig5}(d) shows again the probability of teleportation for constant $\Delta t$ during the respective teleportation times. In this case, all four particles teleport with almost equal probability approaching unity as $\Delta t\rightarrow 0$, and the same would happen for atoms, molecules, even macroscopic objects, with the ``only" added difficulty of sustaining the measurements longer and longer times. For example, to teleport an electron one meter apart one would need about $3\times 10^{-11}$ s and about $2^{20}$ number of measurements each atomic unit of time, i.e., each $24$ attosecond. To teleport a proton requires $1.3\times 10^{-9}$ s and about $2^{26}$ measurements each $24$ attoseconds. For $1$ g, the teleportation time grows up to 16 minutes with $2^{65}$ measurements each $24$ attoseconds, and for 1 kg teleportation takes about 9 hours with about $2^{70}$ each $24$ attoseconds. 

Ultimately, the coherent state of being located at $x_0$ and at about $-x_0$ at the same time will collapse to $x_0$ or $-x_0$ if any interaction with the environment (in addition to interactions involved in the velocity measurements in the QZD) would inform where the particle is located. With increasing environmental interactions of heavier particles and increasing teleportation times, the state will most likely collapse at the very beginning of the teleportation process to the state located at $x_0$.

\section{Conclusions}

We have unveiled a mechanism in which a quantum particle at rest at a classical turning point of a potential, and therefore subjected to a force, remains at rest more likely the more frequently it is measured if it remains at rest. This is an example of QZD which, interestingly, leads to a phenomenon not previously described: Teleportation of the quantum particle, whereby the particle appears and disappears from all points where it can be at rest, i.e., at other turning points of the potential. Disappearing and appearing means here that there is no probability flux exiting the boundaries of the original location nor entering the boundaries of the final location. This entails a violation of the continuity equation of the probability density, which is, in our opinion, the fact that makes this phenomenon to deserve the name teleportation. We have set realistic values of the involved physical quantities showing the increasing difficulty of performing teleportation as the particle gets heavier, but still leaving the door open to perform actual teleportation  experiments, at least for very light particles, for example, with the suggested experimental set up in Fig. \ref{fig1}. Of course, the teleportation of a particle described here must be clearly distinguished from the current studies and experiments of teleportation of quantum information.

As a final remark, we have not considered non-selective measurements in this QZD, as in \cite{porras3,majeed}, because the ``no path" property is lost. While the probability of finding the particle at the other turning point would be higher than with selective measurements \cite{porras3,majeed}, the wave functions corresponding to negative outcomes tend to fill  the path between the two turning points. Thus, QZD-assisted teleportation requires selective measurements.

\acknowledgements

M.A.P. acknowledges support from the Spanish Ministry of Science and Innovation, Gobierno de España, under Contract No. PID2021-122711NB-C21. M.C-A. acknowledges support from grant ``Beca de colaboración de formación" of the Universidad Polit\'ecnica de Madrid.

\end{document}